\documentclass{elsarticle}

\usepackage{epsfig}
\usepackage{a4wide}

\def\as{\alpha_s}
\def\ca{{C^{}_A}}
\def\cf{{C^{}_F}}
\def\nl{{n^{}_{l}}}

\def\lb#1{\if 1#1 \ln\beta \else \ln^#1\beta \fi}
\def\lt#1{\if 1#1 \ln 2 \else \ln^#1 2 \fi}
\def\lsc{L_8}
\newcommand{\bff}[1]{\mbox{\boldmath ${#1}$}}

\begin{document}


 \vspace{\baselineskip}

\title{Threshold expansion of the $gg(q\bar{q})
\rightarrow Q\overline{Q}+X$ cross section at ${\cal O}(\alpha_s^4)$}

    \author[aachen]{Martin Beneke}
    \author[aachen]{Michal Czakon}
    \author[durham]{Pietro Falgari}
    \author[stonybrook]{Alexander Mitov}
    \author[freiburg]{Christian Schwinn}

    \address[aachen]{
      Institut f\"ur Theoretische Physik E,
      RWTH Aachen University,\\  D-52056 Aachen, Germany
    }
    \address[durham]{
      IPPP, Department of Physics, University of Durham, \\
      Durham DH1 3LE, England
    }
    \address[stonybrook]{
      C.N.\ Yang Institute for Theoretical Physics,
      Stony Brook University, \\ Stony Brook, New York 11794--3840, USA
    }
    \address[freiburg]{Albert-Ludwigs Universit\"at Freiburg,
      Physikalisches Institut, D-79104 Freiburg, Germany
    }

\cortext[thanks]{Preprint numbers: PITHA~09/33, 
IPPP/09/93, DCPT/09/186, YITP-SB-09-38}

    \begin{abstract}
      \noindent 
      We derive the complete set of velocity-enhanced terms in the
      expansion of the total cross section for heavy-quark pair production
      in hadronic collisions at next-to-next-to-leading order. 
      Our expression takes into
      account the effects of soft-gluon emission as well as that of
      potential-gluon exchanges. We prove that there are no
      enhancements due to subleading soft-gluon couplings multiplying 
      the leading Coulomb singularity.
    \end{abstract}

    \maketitle

\section{Introduction}

\noindent 
Hadronic heavy-quark pair production is of 
interest not only because of its phenomenological relevance in the 
particular case of top quarks, but also because of
the theoretical insights that may be gained into the singularity 
structure of QCD in the presence of massive partons.

A complete result for the next-to-next-to-leading order (NNLO)
 ${\cal O}(\alpha_s^4)$ corrections is still elusive despite 
substantial recent progress. First, the virtual corrections at NNLO 
were obtained in the limit of large invariants 
$\hat {s},-\hat{t} \gg 4 m^2$ \cite{Czakon:2007ej,Czakon:2007wk}
thanks to the understanding of this limit as an
alternative regularization scheme for collinear divergences
\cite{Mitov:2006xs,Becher:2007cu}. Subsequently, the complete amplitude 
in the case of quark annihilation was derived 
with numerical methods \cite{Czakon:2008zk}, followed by
partial analytic results \cite{Bonciani:2008az,Bonciani:2009nb}.
The latest studies of the structure of massive gauge amplitudes
\cite{Kidonakis:2009ev,Mitov:2009sv,Becher:2009kw,
Beneke:2009rj,Czakon:2009zw,Ferroglia:2009ep,Ferroglia:2009ii} led to
the derivation of the 
soft and collinear divergences of the $gg(q\bar q) \to Q\overline Q$ two-loop
amplitudes in analytical form~\cite{Ferroglia:2009ii}. Despite all
this, the NNLO program for the calculation of the heavy-particle 
pair production cross section will only be completed once 
these two-loop corrections have been combined with squared one-loop
corrections~\cite{Korner:2008bn,Anastasiou:2008vd,Kniehl:2008fd}, the
one-loop corrections to $t\bar t +j$~\cite{Dittmaier:2007wz}, and the real
corrections.

In this note we build on our recent work on soft-gluon radiation 
\cite{Czakon:2009zw} and soft-gluon radiation in the presence 
of Coulomb enhancements~\cite{Beneke:2009rj} to derive the complete 
set of velocity-enhanced terms in the expansion of the total 
heavy-quark pair production cross section at NNLO. The
result contains terms of the form $\beta^i\log^j\beta$, with
$\beta=\sqrt{1-4 m^2/\hat{s}}$ the velocity of the heavy quark, 
and $-2 \leq i \leq 0$, $0 \leq j \leq 2i+4$, which we provide, 
apart from the constant term $i = j = 0$, 
which remains inaccessible with the methods used. As discussed 
in~\cite{Beneke:2009rj}, the $\ln\beta$ coefficient at NNLO 
receives contributions from heavy-quark potentials other than 
the Coulomb potential, and from soft-gluon effects not contained in the 
standard resummation formula for the total cross section due to 
subleading soft-gluon couplings multiplying the leading Coulomb 
singularity. We 
compute these terms below, and prove that the subleading soft-gluon 
effects vanish for the total cross section. Our results may be 
used for improving approximate NNLO top-quark production cross 
section calculations~\cite{Moch:2008qy}. We also provide a 
general formula for the velocity-enhanced terms at NNLO for arbitrary colour 
representations of the particles involved, which requires as only 
process-specific input the constant term of the NLO virtual 
amplitude at threshold in every colour and spin channel.

\section{Sources of enhancement}
\label{sec:sources}

\noindent
There are two general sources of
enhancement of the partonic cross section near threshold. One is
connected with the emission of soft gluons, resulting in 
up to two powers of $\ln\beta$ per emission, whereas the other is due
to potential exchanges of gluons (the Coulomb potential being the
most prominent example), yielding up to one factor of 
$1/\beta$ and $\ln\beta$ per loop.

The NNLL soft-gluon effects can be readily 
derived either by inverting the Mellin-space transform of 
Ref.~\cite{Czakon:2009zw} or, directly in $x$-space, from the 
resummation formula of Ref.~\cite{Beneke:2009rj}. The latter approach 
has been studied
before \cite{Becher:2007ty,Ahrens:2008nc} in the context of the
Drell-Yan and Higgs-boson production processes at hadron colliders.
The soft-gluon enhancement of the Coulomb-potential effects has been
considered at NLO/NLL in Ref.~\cite{Bonciani:1998vc} 
assuming factorization of the two effects. A priori, 
there could be a highly non-trivial
intertwinement between the two. The general factorization of leading
soft-gluon and Coulomb effects has been studied recently 
in Ref.~\cite{Beneke:2009rj} to all orders, resulting in the 
formula
\begin{equation}
\label{eq:fact}
\sigma_{pp'}(\hat s,\mu) = \sum_{i,i'}H_{ii'}(m,\mu)
\;\int d \omega\;
\sum_{R_\alpha}\,J_{R_\alpha}(E-\frac{\omega}{2})\,
W^{R_\alpha}_{ii'}(\omega,\mu)
\end{equation}
for the partonic cross sections, which justifies the multiplicative  
factorization of the two effects in Mellin space to NNLO.  
In the case at hand, where we are only interested in the
NNLO expansion of the resummed result, 
we can describe the problem as a sum of a pure NNLO soft-gluon 
exchange (which requires the knowledge of the two-loop anomalous
dimension \cite{Beneke:2009rj,Czakon:2009zw}) contained in the 
two-loop contribution to the soft function $W^{R_\alpha}_{ii'}$; 
a soft-gluon enhancement of the Coulomb-potential exchange 
due to the convolution of the one-loop terms in $J_{R_\alpha}$ 
and $W^{R_\alpha}_{ii'}$~\cite{Beneke:2009nr}; and, finally, 
two-loop non-Coulomb potential and kinetic-energy corrections 
as described by non-relativistic effective theory (NRQCD), which for 
the present purpose may be thought of as two-loop 
contributions to the non-relativistic function $J_{R_\alpha}$.
By factorizing the ``hard'' cross sections $H_{ii'}(m,\mu)$  from the
soft and Coulomb effects, one generates not only logarithms of the type
$1/\beta\times \log^2\beta$ and $1/\beta\times \log\beta$, but also a
process-dependent non-logarithmic term proportional to $1/\beta$, 
due to the product
of the matching coefficients and the Coulomb potential. Our result 
below differs from the one given 
in Ref.~\cite{Moch:2008qy,Langenfeld:2009wd} 
due to this effect, the value of the two-loop soft 
anomalous dimension \cite{Beneke:2009rj,Czakon:2009zw}, and 
the $\ln\beta$ terms from the non-Coulomb effects, which 
we now derive.

We present two ways to obtain the desired result. The first is based 
on an explicit calculation of the potential contribution in 
NRQCD. We first generalize the expression for the colour-singlet 
heavy-quark potential in momentum space given in Ref.~\cite{Beneke:1999qg} to 
arbitrary colour representations $R_\alpha$. To obtain the 
NNLO $\ln\beta$ terms it is sufficient to use the four-dimensional 
potentials. The required terms read 
\begin{eqnarray}
\tilde{V}(\bff{p},\bff{q}) &=&  
\frac{4\pi D_{R_\alpha}\alpha_s(\mu^2)}{\bff{q}^2} \Bigg[1+ 
\left(a_1-\beta_0\ln\frac{\bff{q}^2}{\mu^2}\right) \frac{\alpha_s}{4\pi} 
\nonumber\\
&&
+\,\frac{\pi\alpha_s(\mu^2)|\bff{q}|}{4 m}
\left(\frac{D_{R_\alpha}}{2}+C_A \right)
+ \frac{\bff{p}^2}{m^2}+\frac{\bff{q}^2}{m^2} \,v_{\rm spin}\Bigg],
\label{delV}
\end{eqnarray}
where $D_{R_\alpha}$ is the strength of the Coulomb potential in 
representation $R_\alpha$, such that $D_{R_\alpha} = -C_F$ 
for the singlet and $D_{R_\alpha} = -(C_F-C_A/2)$ for the 
octet representation (note the sign convention!), and 
$v_{\rm spin} = 0$ and $-2/3$ for a $t\bar t$ pair in a 
spin-singlet and spin-triplet state, respectively.
The non-Coulomb 
potentials, including a new result for the one-loop $1/r^2$-potential 
in an arbitrary representation, are those in the second line. The 
first line refers to the Coulomb potential and its one-loop 
correction that is already dealt with as described above. To 
NNLO we need the resummed insertions of the non-Coulomb potentials 
used in the $e^+ e^-\to t\bar t$ calculation of Ref.~\cite{Beneke:1999qg}, 
given explicitly in Ref.~\cite{Pineda:2006ri}, expanded to 
NNLO, which results in very simple expressions. Including the 
relativistic kinetic-energy correction we 
find the non-Coulomb contribution 
\begin{equation}
\sigma_{X|\rm nC} = \sigma_X^{(0)} 
\,\alpha_s^2(\mu^2)\,\ln\beta \,\left[-2 D_{R_\alpha}^2 \,(1+v_{\rm spin}) 
+ D_{R_\alpha} C_A\right]
\label{eq:non-Coulomb}
\end{equation}
to the total cross section, 
with $ \sigma_X^{(0)}$ the Born cross section in the spin and 
colour channel $X$. For top quarks the Born cross section in the 
$q\bar q$ initiated channel 
is a pure colour-octet spin-triplet, whereas in gluon-gluon fusion 
the $t\bar t$ state is spin-singlet but colour-octet or -singlet.

\begin{figure}[t]
\begin{center}
\includegraphics[width=.5\textwidth]{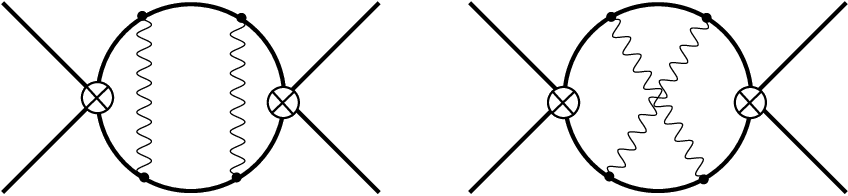}
\end{center}
\caption{\label{fig:color} Graphs relevant to the
  $1/r^2$ potential contributions in the singlet and octet
  channels discussed in the text. Crosses correspond to the singlet or
  octet colour-projection operators.}
\end{figure}

The second derivation of the non-Coulomb logarithms 
uses known results on the threshold expansion 
of the  $e^+e^- \rightarrow t\bar t$
\cite{Czarnecki:1997vz} and $\gamma\gamma\rightarrow t\bar t$
\cite{Czarnecki:2001gi} processes at NNLO. The potential contributions
are implicit in these results, and the two processes cover 
the singlet-triplet spin dependence of the results for the
enhanced terms in exact correspondence to the hadronic case. 
The only non-trivial issue is the colour
dependence
since we also need the colour-octet case. 
It is well known that for an interaction with colour
structure $T^a \otimes T^a$, the transition between singlet and octet
is obtained by a simple change $C_F \rightarrow C_F-C_A/2$
corresponding to the different value of $D_{R_\alpha}$. 
But the $1/r^2$ potential comes also from exchanges of
two gluons as depicted in Fig.~\ref{fig:color}. An explicit check
proves that each of the diagrams gives the correct contribution (as
far as colour is concerned), with the same replacement as before. Thus,
we obtain the correct results for hadronic $t\bar t$ production  
by keeping only the velocity-enhanced terms from the
respective formulae of \cite{Czarnecki:1997vz,Czarnecki:2001gi},
making the replacement $C_F \rightarrow C_F-C_A/2$ for colour-octet
contributions, and removing the contribution from the hard matching
coefficient at one-loop multiplying the one-loop Coulomb
potential. The latter step is crucial in obtaining the correct result,
since the appropriate matching coefficients corresponding to the
processes considered have already been taken into account in the soft gluon
enhancement of the Coulomb contribution as described above.

\begin{figure}[t]
\begin{center}
\includegraphics[width=.5\textwidth]{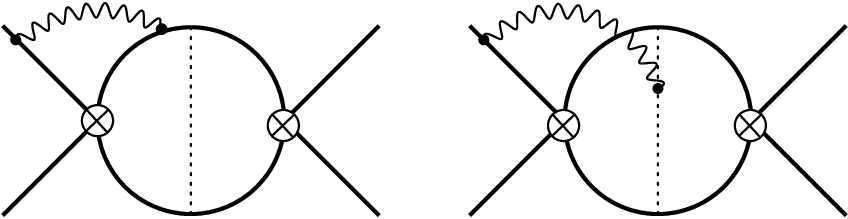}
\end{center}
\caption{\label{fig:mixing} Example graphs with contributions from
both the ultrasoft (gluons depicted with wavy lines) and potential
(gluons depicted with dashed lines) regions. The crosses denote
effective interactions, the structure of which is irrelevant to the
argument of the text.}
\end{figure}

There could be another enhanced single or double 
logarithm of velocity at NNLO from the product of a 
$\alpha_s/\beta$ Coulomb term multiplying an $\alpha_s\beta\ln^2\beta$ 
or $\alpha_s\beta\ln\beta$ term from a beta-suppressed  
subleading soft-gluon coupling~\cite{Beneke:2009rj}. 
Such suppressed couplings exist for the emission of soft 
gluons from the initial state as well as from the final state. 
We now show that such terms do not appear in the total pair 
production cross section.\footnote{This can be anticipated
from the known expansion of the NLO cross section
\cite{Czakon:2008ii}. From this result one can readily verify that
the logarithms of velocity appear in terms suppressed by even powers
of $\beta$ relative to the leading terms, i.e. no 
$\alpha_s \beta\log\beta$ terms of the mentioned origin are
generated.} To this end we imagine obtaining the cross section 
by evaluating the imaginary part of forward-scattering graphs 
such as those of Fig.~\ref{fig:mixing}. 
We first consider the subleading coupling to the heavy-quark loop, 
so the gluon coupling to the external line in
Fig.~\ref{fig:mixing} is the standard eikonal coupling. In 
the framework of non-relativistic effective theory the subleading 
gluon coupling corresponds to the $\bff{x}\cdot \bff{E}$ 
interaction~\cite{Voloshin:1978hc,Pineda:1997bj}.
An expansion of the heavy-quark loop in the velocity can be extracted
directly by the strategy of regions \cite{Beneke:1997zp}. 
As described in the latter work,
it is sufficient to consider the following regions of 
integration momenta in the partonic cms frame where
the sum of the heavy-particle momenta is $(2m, \vec 0)$: 
hard ($k \sim m$, with $m$ the heavy-quark mass 
and $k$ a loop momentum), potential ($k^0 \sim m
\beta^2$, ${\bf k} \sim m\beta$), soft ($k^0 \sim m \beta$, ${\bf k}
\sim m\beta$) and ultrasoft ($k^0 \sim m \beta^2$, ${\bf k} \sim
m\beta^2$). At NNLO the diagrams
corresponding to the soft region vanish, as they generate only
scaleless integrals. The source of singular terms is in the
potential and ultrasoft regions. By the velocity scaling only
terms corresponding to the potential three-momentum contribute odd
powers of $\beta$. The same scaling arguments also show that, in any
denominator containing a combination of a potential and an ultrasoft
momentum, the ultrasoft momentum will be (multipole) expanded. 
Therefore, the
denominators containing potential three-momenta will not depend on the
direction of any external three-momentum (unlike denominators containing
an ultrasoft three-momentum). In consequence, rotational invariance 
implies that all integrals with an odd number of potential
three-momenta in the
numerator vanish. Thus, given a term with a specified power of
$\beta$, the next higher-order contribution will be suppressed by a
relative factor of $\beta^2$, smaller than the terms we seek.

Next, regarding the subleading 
soft-gluon couplings to the initial state, the relevant expansion 
is one in transverse momentum, since the collinear momenta scale as
$n_+ k \! \sim \! m,n_- k \! \sim \! m\beta^2, k_\perp \! \sim \!
m\beta$. The effective Lagrangian
for the corrections to the eikonal approximation is given in soft-collinear 
effective theory by $\bar{\xi} \left( x_\perp^\mu n_-^\nu \, W_c 
\,gF_{\mu\nu}^{\rm 
us}W_c^\dagger \right) \frac{\not\!n_+}{2} \, \xi$ 
for quarks \cite{Beneke:2002ph,Beneke:2002ni}, and similar terms 
involving transverse derivatives or factors of $x_\perp$ for 
the couplings to collinear gluons, and of soft quarks. None of 
these terms can contribute a beta-suppressed term, since the 
initial-state momenta in Fig.~\ref{fig:mixing} can always be chosen 
to have zero transverse momentum, implying that loop integrals 
with transverse-momentum factors in the numerator vanish by 
arguments similar to those applied to the heavy-quark couplings. 
This completes the proof, that we have correctly taken
into account all possible sources of singular terms in the expansion
of the cross sections for heavy-quark pair production at NNLO by 
including the extra terms from the non-Coulomb potentials. 

Note that some of the cuts of Fig.~\ref{fig:mixing} correspond to 
three-particle colour correlations at the amplitude level, for which 
the infrared divergence structure has recently been given in 
Ref.~\cite{Ferroglia:2009ii}. The latter work shows that the 
infrared-singular three-particle 
correlations may not vanish in the limit $\beta\to 0$ in the 
amplitude, but that they do in the virtual contributions to the total 
cross section at NNLO in the particular case of top 
quarks because of colour projections
\cite{Czakon:2009zw,Ferroglia:2009ii}. Our arguments above prove 
that there are no contributions to the $\ln\beta$ terms 
from three-particle correlations in both, the virtual and real 
corrections. This holds independent of particular colour
representations for purely kinematic reasons.

\section{Results}

\noindent
Next we present the main result of this paper, namely the expansion
of the two-loop partonic cross section close to the partonic
threshold $\beta=0$. As we emphasized above, our result is complete
up to the so-called constant terms\footnote{This standard 
terminology is somewhat misleading in this
process. Due to the non-trivial $\beta$ dependence of the Born
cross section, the contribution of the ``constant" term to the
cross section is, in fact, proportional to $\beta$.} 
$ C^{(2)}_{q\bar q},C^{(2)}_{gg,\bf 1},C^{(2)}_{gg,\bf 8}$. 
Their derivation requires a dedicated calculation that goes beyond the
scope of the present work.
Setting $\mu_R=\mu_F=\mu$, the result for the total cross-section
close to threshold reads:
\begin{eqnarray}
\sigma_{ij,\bf I}(\beta,\mu,m) &=& \sigma^{(0)}_{ij,\bf I} \Bigg\{ 1
+ \frac{\as(\mu^2)}{4\pi} \left[\sigma^{(1,0)}_{ij,\bf I} +
\sigma^{(1,1)}_{ij,\bf I}\ln\left({\mu^2\over m^2}\right) \right]
\label{eq:result-scales}\\
&&+\left(\frac{\as(\mu^2)}{4\pi}\right)^2
\left[\sigma^{(2,0)}_{ij,\bf I} + \sigma^{(2,1)}_{ij,\bf
I}\ln\left({\mu^2\over m^2}\right) + \sigma^{(2,2)}_{ij,\bf
I}\ln^2\left({\mu^2\over m^2}\right)\right] + {\cal O}(\as^3)
\Bigg\} \; ,\nonumber
\end{eqnarray}
where ${\bf I} = {\bf 1},{\bf 8}$ is a colour index and
$ij=(q\bar{q},gg)$, whereas $\alpha_s(\mu^2)$ is defined in the
$\overline{\mbox{MS}}$ scheme with $n_l$ (number of massless quarks)
flavours. The derivation of the coefficients of 
$\ln^n(\mu^2/m^2)$ with $n=1,2$ from one-loop results and 
splitting functions is given in \ref{app:scales}.
The non-trivial scale-independent 
two-loop contributions $\sigma^{(2,0)}_{i j, \bf I}$
read:\footnote{For Eq.~(\ref{eq:sigma-2-0-qq}), see erratum attached.}
\begin{eqnarray}
\sigma^{(2,0)}_{q\bar{q},\bf 8} &=&
\frac{(2\cf-\ca)^2\pi^4}{3\beta^2}+\frac{(2\cf-\ca)\pi^2}{9\beta}
\Big[288\cf\lb2+6\big(48\cf\lt1-23\ca+2\nl\big)\lb1 \nonumber \\ &&
+12\cf\big(-24+9\lt1+\pi^2\big)+3\ca\big(89-58\lt1-3\pi^2\big)+6\nl\big(-5
+6\lt1\big)-32\Big] \nonumber \\ &&
+512C_F^2\lb4+\frac{128}{9}\cf\Big[72\cf\big(-2+3\lt1\big)-29\ca+2\nl\Big]\lb3
\nonumber \\ &&
+\frac{16}{9}\Big[2\cf\big(12\cf(120-207\lt1+156\lt2-7\pi^2)+3\ca(217-198\lt1
-4\pi^2) \nonumber \\ &&
+6\nl(-9+10\lt1)-32\big)+3\ca(17\ca-2\nl)\Big]\lb2 \nonumber \\ &&
+\frac{8}{27}\Big[2\cf\big(18\cf(-960+\lt1(1368-84\pi^2)-1140\lt2+576\lt3
+55\pi^2+336\zeta_3) \nonumber \\[0.1cm] &&
+\ca(-7582+108\lt1(115-2\pi^2)-5886\lt2+360\pi^2+189\zeta_3)
\nonumber \\[0.2cm] && +2\nl(338-630\lt1+378\lt2-9\pi^2)+192(2-3\lt1)\big)
\nonumber \\[0.1cm] &&
+3\ca\big(3\ca(-185+126\lt1+6\pi^2-6\zeta_3)+6\nl(11-10\lt1)+32\big)\Big]\lb1
 + C^{(2)}_{q\bar q} \; , \label{eq:sigma-2-0-qq}
\\[0.5cm] \nonumber 
\sigma^{(2,0)}_{gg,\bf 1} &=& \frac{4C_F^2
\pi^4}{3\beta^2}+\frac{2\cf\pi^2}{9\beta}\Big[288\ca\lb2
+6\big(\ca(-11+48\lt1)+2\nl\big)\lb1 \nonumber \\ &&
+9\cf\big(-20+\pi^2\big)+\ca\big(67-66\lt1+3\pi^2\big)+2\nl\big(-5
+6\lt1\big)\Big]+512C_A^2\lb4 \nonumber \\ &&
+\frac{128}{9}\ca\Big[\ca\big(-155+216\lt1\big)+2\nl\Big]\lb3
+\frac{32}{9}\ca\Big[9\cf\big(-20+\pi^2\big) \nonumber \\ &&
+\ca\big(1963-2790\lt1+1872\lt2-96\pi^2\big)+2\nl\big(-17+18\lt1\big)\Big]\lb2
\nonumber \\ &&
+\frac{16}{27}\Big[27\cf\big(-2\cf\pi^2+\ca(80+6\lt1(-20+\pi^2)-5\pi^2)\big)
+\ca\big(\ca(-23758 \nonumber \\[0.2cm] &&
+18\lt1(1963-96\pi^2)-24246\lt2+10368\lt3+1251\pi^2+6237\zeta_3)
\nonumber \\[0.2cm] && +2\nl(218-306\lt1+162\lt2-9\pi^2)\big)\Big]\lb1
 + C^{(2)}_{gg,\bf 1} \; , \label{eq:sigma-2-0-gg1}
\\[0.5cm] \nonumber 
\sigma^{(2,0)}_{gg,\bf 8} &=&
\frac{(2\cf-\ca)^2\pi^4}{3\beta^2}+\frac{(2\cf-\ca)\pi^2}{18\beta}
\Big[576\ca\lb2+12\big(\ca(-23+48\lt1)+2\nl\big)\lb1 \nonumber \\ &&
+18\cf\big(-20+\pi^2\big)+\ca\big(278-132\lt1-3\pi^2\big)+4\nl\big(-5
+6\lt1\big)\Big] +512C_A^2\lb4 \nonumber \\ &&
+\frac{128}{9}\ca\Big[\ca\big(-173+216\lt1\big)+2\nl\Big]\lb3+\frac{16}{9}\ca
\Big[18\cf\big(-20+\pi^2\big) \nonumber \\ &&
+\ca\big(4553-6156\lt1+3744\lt2-201\pi^2\big)+2\nl\big(-37+36\lt1\big)\Big]\lb2
\nonumber \\ &&
+\frac{4}{27}\Big[54\cf\big(-4\cf\pi^2+\ca(180+12\lt1(-20+\pi^2)-7\pi^2)\big)
+\ca\big(\ca(-111418 \nonumber \\[0.2cm] &&
+36\lt1(4499-201\pi^2)-105624\lt2+41472\lt3+5823\pi^2+24840\zeta_3)
\nonumber \\[0.2cm] && +4\nl(505-666\lt1+324\lt2-18\pi^2)\big)\Big]\lb1
 + C^{(2)}_{gg,\bf 8} \; . \label{eq:sigma-2-0-gg8}
\end{eqnarray}

In order to construct the two-loop correction to the colour-averaged
cross section from the colour-state specific components
given above, one has to first multiply the two-loop contributions
Eqs.~(\ref{eq:sigma-2-0-qq},\ref{eq:sigma-2-0-gg8}) and
(\ref{eq:sigma-2-0-gg1}) by, respectively, $\sigma^{(0)}_{ij,\bf 8}$
and $\sigma^{(0)}_{ij,\bf 1}$ (see Eq.~(\ref{eq:result-scales})),
and then add them together. The singlet/octet Born terms can be
found in \ref{app:scales}. Finally, by setting $\mu=m$,
all colour factors to their numerical values, and $n_l=5$ as 
applicable to top-quark production, we get
the following result for the colour-averaged total inclusive
cross-section close to partonic threshold:\footnote{For 
the first equation, see erratum attached.}
\begin{eqnarray}
\sigma^{(2)}_{q\bar{q}} &=&
\frac{3.60774}{\beta^2}
+\frac{1}{\beta}\Big(-140.368\lb2+32.106\lb1+3.95105\Big) \nonumber \\ &&
+910.222\lb4-1315.53\lb3+592.292\lb2+528.557\lb1+C^{(2)}_{qq} \; ,
\nonumber \\[0.5cm]
\sigma^{(2)}_{gg} &=&
\frac{68.5471}{\beta^2}
+\frac{1}{\beta}\Big(496.3\lb2+321.137\lb1-8.62261\Big) \nonumber \\ &&
+4608\lb4-1894.91\lb3-912.349\lb2+2456.74\lb1+C^{(2)}_{gg} \; ,
\end{eqnarray}
which differs in the coefficients of the  $1/\beta$ and $\ln\beta$ 
terms from the expressions given in~\cite{Langenfeld:2009wd} 
for the reasons mentioned in section~\ref{sec:sources}.

\vskip0.2cm
\noindent
In conclusion, the above formulae contain all velocity-enhanced terms 
in the total hadronic production of heavy quarks at NNLO 
near the partonic threshold. A compact general result for the 
velocity-enhanced terms in the 
production of equal-mass heavy-particle pairs in the collisions 
of massless particles for arbitrary colour representations is 
provided in appendix~A. 

\noindent
\section*{Acknowledgments}

\noindent 
A.M. would like to thank K.~Melnikov and G.~Sterman for very insightful
discussions. The work of M.B. is supported by the 
DFG Sonder\-forschungsbereich/Transregio~9 
``Computergest\"utzte Theoretische Teilchenphysik''. 
The work of M.C. is supported by the Heisenberg and by the Gottfried
Wilhelm Leibniz programmes of the Deutsche Forschungsgemeinschaft. The
work of A.M. is supported by a fellowship from the {\it US LHC Theory
  Initiative} through NSF grant PHY-0705682 as well as by NSF grant
PHY-0653342. The work of P.F. is supported in part by the grant
``Premio Morelli-Rotary 2009'' of the Rotary Club Bergamo.

\appendix

\section{A formula for arbitrary representations}

\noindent
Here we provide the velocity-enhanced terms at NNLO for the 
production of a pair of heavy particles with equal mass $m$ 
in the scattering 
of massless partons in colour representations $r$ and $r'$, 
respectively, under the assumption that the Born cross section 
(which is factored out as in Eq.~(\ref{eq:result-scales})) 
admits an $S$-wave term proportional to $\beta$.
The heavy-particle pair is in colour representation 
$R_\alpha$ and a definite spin state. The threshold expansion 
reads 
\begin{eqnarray}
\sigma^{(2)}_X &\!\!=\!\!& 
\frac{4 \pi^4 D_{R_\alpha}^2}{3 \beta^2} +
\frac{\pi^2 D_{R_\alpha}}{\beta} \,
\bigg\{(-8) \,(C_r+C_{r'}) \bigg[\ln^2\left(\frac{2 m\beta^2}{\mu}\right)
-\frac{\pi^2}{8}\bigg]
+ 2 \,(\beta_0+4  C_{R_\alpha}) \ln\left(\frac{2 m\beta^2}{\mu}\right)
\nonumber\\
&& \hspace*{1cm} 
- \,8  C_{R_\alpha}-2 a_1 -4 \,\mbox{Re}\,[C^{(1)}_X] 
+2 \beta_0  \ln\left(\frac{2 m}{\mu}\right)
\bigg\}
\nonumber\\
&& 
+\, 128 \,(C_r+C_{r'})^2 \ln^4 \beta 
+ 64 \,(C_r+C_{r'}) \left\{4 \,(C_r+C_{r'}) 
\left(\lsc-2\right) -\frac{\beta_0}{3} -2 C_{R_\alpha}\right\} 
\ln^3\beta
\nonumber\\
&&+\,\bigg\{ \frac{8}{3} \,(C_r+C_{r'})^2 
\left[72  \lsc^2   -288  \lsc + 576-35 \pi^2\right]
+\frac{16}{9} \,(C_r+C_{r'}) \,\Big[18 \,{\rm Re}\,[C^{(1)}_X] 
\nonumber\\
&& \hspace*{0.6cm} 
+\, 18 \beta_0 \left(-\lsc+2\right) 
+36 C_{R_\alpha} \left(-3\lsc+7\right)+C_A (67-3 \pi^2)-20 n_l T_f
\Big]
\nonumber\\
&&\hspace*{0.6cm} 
+\,16 C_{R_\alpha} (\beta_0+2 C_{R_\alpha})
\phantom{\hspace{-0.2cm}\frac{|}{|}}\bigg\} \ln^2 \beta 
\nonumber\\
&&
+ \,\bigg\{ 8 \,(C_r+C_{r'})^2 \left[
8 \lsc^3 -48\lsc^2 +\left(192-\frac{35 \pi^2}{3}\right)
\lsc -384+\frac{70 \pi^2}{3}+112 \zeta_3 \right] 
\nonumber\\
&&\hspace*{0.6cm}
+\,2 \,(C_r+C_{r'}) \left[-16 \,\mbox{Re}\,[C^{(1)}_X] \left(-\lsc+2\right) 
+\beta_0 \left(-8\lsc^2+32 \lsc -64+\frac{11 \pi^2}{3}\right)\right.  
\nonumber\\
&&\hspace*{1cm} 
+\,2 C_{R_\alpha} \left(-24 \lsc^2+112 \lsc -224 
+\frac{35\pi^2}{3}\right)
\nonumber\\
&&\hspace*{1cm}  
+ \, C_A \left(\frac{8}{3} \left(\frac{67}{3}-\pi^2\right)\lsc
  -\frac{4024}{27}+\frac{59 \pi^2}{9}+28 \zeta_3\right)
\left.  +\frac{4 n_l T_f}{9} \left(\!-40\lsc+\frac{296}{3}-\pi^2\right)\right]
\nonumber\\
&&\hspace*{0.6cm}
+\,4 \,C_{R_\alpha} 
\bigg[ -4 \,\mbox{Re}\,[C^{(1)}_X] 
-4 \,(\beta_0+2 C_{R_\alpha}) \left(-\lsc+3\right)
+C_A \left(-\frac{98}{9}+\frac{2 \pi^2}{3}-4 \zeta_3
    \right)
\nonumber\\
&&\hspace*{1cm}
+\frac{40}{9} n_l T_f \bigg]
+16 \pi^2 D_{R_\alpha} \Big[C_A-2 D_{R_\alpha} (1+v_{\rm spin}) \Big] 
\bigg\} \,\ln \beta+{\cal O}(1)\,. 
\label{eq:general}
\end{eqnarray}
We obtained this result by expanding the resummation 
formula (\ref{eq:fact}) to NNLO and 
adding the non-Coulomb terms according to Eq.~(\ref{eq:non-Coulomb}).
$C_r$, $C_{r'}$ and $C_{R_\alpha}$ denote the quadratic Casimir
operators of the colour representations, $\beta_0 =
\frac{11}{3} C_A-\frac{4}{3} n_l T_f$ is the one-loop beta-function 
coefficient, and $L_8 = \ln(8 m/\mu)$. The quantities $D_{R_\alpha}$, 
$a_1 =\frac{31}{9} C_A-\frac{20}{9} n_l T_f$ and $v_{\rm spin}$ 
are connected with the heavy-quark potentials as discussed in the 
main text. $C^{(1)}_X$ is the one-loop hard matching coefficient 
in the resummation formula for hadronic heavy-particle 
pair production at threshold~\cite{Beneke:2009rj}, when the 
heavy-particle pair is in colour and spin state $X$. Alternatively, 
it can be deduced from the constant term in the 
threshold limit of the NLO 
production cross section $\sigma^{(1)}_X$ 
in colour and spin channel $X$ by comparing the 
expansion  of $\sigma^{(1)}_X$ to the formula 
\begin{eqnarray}
\sigma^{(1)}_X &\!\!=\!\!& 
- \frac{2 \pi^2 D_{R_\alpha}}{\beta} +
4\,(C_r+C_{r'}) \bigg[\ln^2\left(\frac{8 m\beta^2}{\mu}\right)
+8 -\frac{11 \pi^2}{24}\bigg]
\nonumber\\
&&  
- \,4 \,(C_{R_\alpha}+ 4\,(C_r+C_{r'})) \,
\ln\left(\frac{8 m\beta^2}{\mu}\right) 
+ 12  C_{R_\alpha} +2 \,\mbox{Re}\,[C^{(1)}_X] + 
{\cal O}(\beta)\,.
\label{eq:NLOexp}
\end{eqnarray}
The results for $t\bar t$ production in the main text 
can be generated from the general formula by inserting the relevant 
colour and spin factors. The required matching coefficients  
$\mbox{Re}\,C^{(1)}_X$ can be determined from 
\cite{Hagiwara:2008df,Czakon:2008cx} (for convenience of the reader,
the NLO cross sections are reproduced in Eq.~(\ref{eq:nlosigma}) 
below) and (\ref{eq:NLOexp}).  
Note that the cubic $\ln\mu$ dependence in the $L_8^3$ term 
in (\ref{eq:general}) cancels with a corresponding term in the 
product $\mbox{Re}\,[C^{(1)}_X]\,L_8$ as required since 
$\sigma^{(2)}_X$ can depend on $\ln\mu$ at most quadratically.

\section{Derivation of the scale dependence}
\label{app:scales}

\noindent
The scale-dependent terms in Eq.~(\ref{eq:result-scales}) can 
be obtained from Eq.~(\ref{eq:general}). Here we give an 
independent derivation from the known LO and NLO results; 
the procedure has been
detailed, for example, in Ref.~\cite{Anastasiou:2002yz} and we adopt
it in the following. To the best of our knowledge, these terms are
not available in the literature in completely analytical form.

We consider the case $\mu_R=\mu_F=\mu$. To simplify the
following discussion, we introduce the functions $s^{(a,b)}_{ij,\bf
I} = 2^{-a}\beta\,\sigma^{(a,b)}_{ij,\bf I}$ with $s^{(0,0)}_{ij,\bf
I} = \beta$. We then rewrite Eq.~(\ref{eq:result-scales}) as:
\begin{eqnarray}
\sigma_{ij,\bf I}(\beta,\mu,m) &=& {\sigma^{(0)}_{ij,\bf I} \over
\beta} \Bigg\{ \beta + \frac{\as(\mu^2)}{2\pi}
\left[s^{(1,0)}_{ij,\bf I} + s^{(1,1)}_{ij,\bf
I}\ln\left({\mu^2\over m^2}\right) \right]
\label{eq:result-scales-app}\\
&&+\left(\frac{\as(\mu^2)}{2\pi}\right)^2 \left[s^{(2,0)}_{ij,\bf I}
+ s^{(2,1)}_{ij,\bf I}\ln\left({\mu^2\over m^2}\right) +
s^{(2,2)}_{ij,\bf I}\ln^2\left({\mu^2\over m^2}\right)\right] +
{\cal O}(\as^3) \Bigg\} \; .\nonumber
\end{eqnarray}
Note that as follows from Eq.~(\ref{eq:Born}) below, the overall
factor ${\sigma^{(0)}_{ij,\bf I}/\beta}$ is a beta independent 
constant. It is factored out for later convenience. We will
not consider the $ij=qg$ subprocess since $\sigma_{qg,\bf I} = {\cal
O}(\beta^3)$. The scaling functions appearing in
Eq.~(\ref{eq:result-scales-app}) read:
\begin{eqnarray}
\hat s^{(1,1)}_{ij,\bf I}(\rho) &=& \beta_0\hat s^{(0,0)}_{ij,\bf
I}(\rho) - P^{(0)}_{ik} \otimes \hat s^{(0,0)}_{kj,\bf I} (\rho) -
\hat s^{(0,0)}_{ik,\bf I}
\otimes P^{(0)}_{kj} (\rho) \; ,
\label{eq:G1hat}
\nonumber \\
\hat s^{(2,2)}_{ij,\bf I}(\rho) &=& {3\over 4}\beta_0\hat
s^{(1,1)}_{ij,\bf I}(\rho) - {1\over 2} P^{(0)}_{ik} \otimes \hat
s^{(1,1)}_{kj,\bf I} (\rho) -
{1\over 2} \hat s^{(1,1)}_{ik,\bf I} \otimes P^{(0)}_{kj} (\rho) \; , 
\nonumber \\
\hat s^{(2,1)}_{ij,\bf I}(\rho) &=& {3\over 2}\beta_0\hat s^{(1,0)}_{ij,\bf
I}(\rho) + {1\over 2}\beta_1\hat s^{(0,0)}_{ij,\bf I}(\rho) - P^{(0)}_{ik}
\otimes
\hat s^{(1,0)}_{kj,\bf I} (\rho) \nonumber\\
&& - \hat s^{(1,0)}_{ik,\bf I} \otimes P^{(0)}_{kj} (\rho) -
P^{(1)}_{ik} \otimes \hat s^{(0,0)}_{kj,\bf I} (\rho) - \hat
s^{(0,0)}_{ik,\bf I} \otimes P^{(1)}_{kj} (\rho) \; .
\label{eq:H2hat}
\end{eqnarray}
In the above equation we need to consider the cross sections as
functions of the dimensionless variable $\rho=4m^2/\hat s$, and with
$\beta=\sqrt{1-\rho}$. We have also introduced the ``hat'' notation
such that for any function $\sigma(\rho)$ we have $\hat\sigma(\rho)
\equiv \sigma(\rho)/\rho$; for its origin see the discussion in
Ref.~\cite{Anastasiou:2002yz}. Also, summation over repeated indexes
$k=q,\bar{q},g$ is understood and $\beta_0$ and $\beta_1$ are the
QCD beta function coefficients
\begin{eqnarray}
\label{eq:betafct} \beta_0 = {11 \over 3}\ca - {2 \over 3}\nl \, ,
\qquad \beta_1 = {34 \over 3}C_A^2 - {10 \over 3}\ca\nl - 2\cf\nl \, ,
\end{eqnarray}
with $\ca = 3$ and $\cf = 4/3$. Throughout this letter we take the
number of active quarks to be equal to the number of light quarks
$\nl$.
The functions $P^{(n)}$ are the one- and two-loop space-like
DGLAP evolution kernels defined as an expansion in $\as/(2\pi)$.

For the reactions of interest, and neglecting terms that contribute
beyond ${\cal O}(\beta)$, Eqs.~(\ref{eq:H2hat}) 
reduce to:
\begin{eqnarray}
\hat s^{(1,1)}_{q\bar q,\bf I}(\rho) &=& \beta_0\hat
s^{(0,0)}_{q\bar q,\bf I}(\rho) - 2P^{(0)}_{qq}
\otimes \hat s^{(0,0)}_{q\bar q,\bf I} (\rho)\; ,
\nonumber \\
\hat s^{(2,2)}_{q\bar q,\bf I}(\rho) &=& {3\over 4}\beta_0\hat
s^{(1,1)}_{q\bar q,\bf I}(\rho) - P^{(0)}_{qq}
\otimes \hat s^{(1,1)}_{q\bar q,\bf I} (\rho) + {\cal O}(\beta^3) \; ,
\nonumber \\
\hat s^{(2,1)}_{q\bar q,\bf I}(\rho) &=& {3\over 2}\beta_0\hat
s^{(1,0)}_{q\bar q,\bf I}(\rho) + {1\over 2}\beta_1\hat s^{(0,0)}_{q\bar
q,\bf I}(\rho) - 2P^{(0)}_{qq} \otimes \hat s^{(1,0)}_{q\bar q,\bf
I}(\rho) - 2P^{(1)}_{qq} \otimes \hat s^{(0,0)}_{q\bar q,\bf I}
(\rho) +{\cal O}(\beta^3) \; . \qquad \label{eq:H2hat-explicit}
\end{eqnarray}
The results for the $gg$ initiated subprocess can be obtained from
Eqs.~(\ref{eq:H2hat-explicit})
by replacing everywhere the pairs $qq$ and $q\bar q$ by $gg$, 
see also Ref.~\cite{Kidonakis:2001nj}.

The above convolutions can be performed in a straightforward manner,
by expanding the splitting functions around $x=1$ and keeping only
the $1/[1-x]_+$ terms and $\delta$-functions. To make our
presentation self contained, in the following we present all terms
from Eq.~(\ref{eq:result-scales}) that are not already given in
Eqs.~(\ref{eq:sigma-2-0-qq},\ref{eq:sigma-2-0-gg1},\ref{eq:sigma-2-0-gg8}).
We do not present the terms ${\cal O}(1)$, i.e.~the terms that are
not enhanced by inverse power of $\beta$ or by a power of $\lb1$,
since they are at the same level as the presently unknown
coefficients $C^{(2)}_{q\bar q},C^{(2)}_{gg,\bf 1}$ and
$C^{(2)}_{gg,\bf 8}$ appearing in
Eqs.~(\ref{eq:sigma-2-0-qq},\ref{eq:sigma-2-0-gg1},\ref{eq:sigma-2-0-gg8}).
Nevertheless, we have calculated these ${\cal O}(1)$ terms and they
can be found in electronic form available with the preprint of this
paper.

The leading-order (Born) terms appearing in
Eq.~(\ref{eq:result-scales}) read:
\begin{eqnarray}
\sigma^{(0)}_{q\bar q,\bf 8} &=& \pi\beta{(N^2-1)\over 8N^2}
{\as^2(\mu^2)\over m^2}\, ,\nonumber\\
\sigma^{(0)}_{gg,\bf 8} &=& \pi\beta{(N^2-4)\over 8N(N^2-1)}
{\as^2(\mu^2)\over m^2} \, ,\nonumber\\
\sigma^{(0)}_{gg,\bf 1} &=& \pi\beta{1\over 4N(N^2-1)}
{\as^2(\mu^2)\over m^2} \, \label{eq:Born}
\end{eqnarray}
with $N=3$ the number of colours.
The NLO terms appearing in Eq.~(\ref{eq:result-scales}) read:
\begin{eqnarray}
\sigma^{(1,1)}_{q\bar q,\bf 8} &=& \frac{22}{3}\ca -\frac{4}{3}\nl
+\cf(10-16\lt1)-16\cf\lb1 \, , 
\nonumber \\
\sigma^{(1,0)}_{q\bar q,\bf 8} &=& (2\cf-\ca)\frac{\pi^2}{\beta} +
32\cf\lb2 + \left[32\cf(-2+3\lt1) -8\ca\right]\lb1 \nonumber\\
&&-\frac{4}{3}\cf(-24+\pi^2+63\lt1-48\lt2)  + \frac{1}{9}\ca(308 -
9\pi^2 - 180\lt1) \nonumber\\
&& + \frac{4}{9}\nl(-5 + 6\lt1) -\frac{32}{9} \, ,
\nonumber\\
\sigma^{(1,1)}_{gg,\bf 8} &=& -16\ca(-1+\lt1)-16\ca\lb1 \, , 
\nonumber \\
\sigma^{(1,0)}_{gg,\bf 8} &=& (2\cf-\ca)\frac{\pi^2}{\beta} +
32\ca\lb2 + 24\ca(-3 + 4\lt1)\lb1 \nonumber\\
&&+\cf(-20+\pi^2)+\frac{1}{6}\ca (504-17\pi^2-624\lt1+384\lt2)\, ,
\nonumber \\
\sigma^{(1,1)}_{gg,\bf 1} &=& \sigma^{(1,1)}_{gg,\bf 8} \, , 
\nonumber \\
\sigma^{(1,0)}_{gg,\bf 1} &=& 2\cf\frac{\pi^2}{\beta} +
32\ca\lb2 + 32\ca(-2 + 3\lt1)\lb1 \nonumber\\
&& + \cf(-20+\pi^2)+\frac{1}{3}\ca (204-7\pi^2-288\lt1+192\lt2) \, .
\label{eq:nlosigma}
\end{eqnarray}

Finally, the NNLO terms in Eq.~(\ref{eq:result-scales}) proportional
to $\ln^n(\mu^2/m^2)$ read:
\begin{eqnarray}
\sigma^{(2,2)}_{q\bar q,\bf 8} &=& 128\cf^2\lb2
+\left[32\cf^2 (-5+8\lt1) -\frac{440}{3}\cf\ca +
\frac{80}{3}\cf\nl \right]\lb1 + {\cal O}(1) \, , 
\nonumber \\
\sigma^{(2,1)}_{q\bar q,\bf 8} &=& \frac{\pi^2}{\beta}\left[
(16\ca\cf-32\cf^2)\lb1 -11\ca^2-12\cf^2-4\cf\nl+\ca(28\cf+2\nl)
\right] 
\nonumber \\
&&-512\cf^2\lb3 + \left[480\cf\ca - 64\cf\nl - 64\cf^2(-21 + 32
\lt1)\right]\lb2 \nonumber\\
&&+\left[ -88\ca^2+16\ca\nl+\frac{64}{3}\cf^2(-102+7\pi^2+156
\lt1-120\lt2) \right.\nonumber\\
&&\left.+\cf\left(\frac{512}{9}-\frac{32}{3}\nl(-17+22\lt1)+\frac{32}{3}
\ca(-136+2\pi^2+141\lt1)\right) \right]\lb1 + {\cal O}(1)\, , \nonumber\\
\sigma^{(2,2)}_{gg,\bf 8} &=& 128\ca^2\lb2
+\left[\frac{8}{3}\ca^2(-107+96\lt1)
+ \frac{16}{3}\ca\nl \right]\lb1 + {\cal O}(1) \, , 
\nonumber \\
\sigma^{(2,1)}_{gg,\bf 8} &=& \frac{\pi^2}{\beta}\left[(16\ca^2-32
\ca\cf)\lb1 -\frac{11}{3}\ca^2+\cf\left(\frac{22}{3}\ca -
\frac{4}{3}\nl\right)+\frac{2}{3}\ca\nl \right] 
\nonumber\\
&&-512\ca^2\lb3 +\left[ - \frac{32}{3}\ca^2(-167 + 192\lt1)
-\frac{64}{3}\ca\nl \right]\lb2 \nonumber\\
&&+\left[ -16\cf\ca(-20 + \pi^2) - \frac{16}{9}\ca\nl(-37 + 36
\lt1)\right.\nonumber\\
&&+\left. \frac{8}{9}\ca^2(-4391 + 201\pi^2 + 5292\lt1 - 2880\lt2)
\right]\lb1 + {\cal O}(1) \, ,
\nonumber\\
\sigma^{(2,2)}_{gg,\bf 1} &=& \sigma^{(2,2)}_{gg,\bf 8} \, , 
\nonumber\\
\sigma^{(2,1)}_{gg,\bf 1} &=& \frac{\pi^2}{\beta}\left[-32\cf\ca\lb1
+\frac{22}{3}\cf\ca - \frac{4}{3}\cf\nl \right] -512\ca^2\lb3
+\left[-\frac{64}{3}\ca\nl \right.\nonumber\\
&&\left. + \frac{32}{3}\ca^2(155 - 192\lt1)\right]\lb2 +\left[
16\cf\ca(20-\pi^2) +\frac{32}{9}\ca\nl(17-18\lt1) \right.
\nonumber\\
&&\left.+\frac{16}{9} \ca^2(-1963+96\pi^2+2502\lt1-1440\lt2)
\right]\lb1 + {\cal O}(1) \, .
\label{eq:nnlosclae}
\end{eqnarray}
Eqs.~(\ref{eq:nnlosclae}) are in agreement with the partially
numerical results provided in Ref.~\cite{Langenfeld:2009wd}.


\newpage

\vspace{\baselineskip}

\begin{center}
\Large
Erratum to \\
``Threshold expansion of the $gg(q\bar{q}) \rightarrow Q\overline{Q}+X$ \\
cross section at ${\cal O}(\alpha_s^4)$'' \\
$[$ Phys.\ Lett.\ B690 (2010) 483-490 $]$
\end{center}
\vskip0.3cm

\centerline{Martin Beneke${}^a$, Michal Czakon${}^b$, Pietro Falgari${}^c$, 
Alexander Mitov${}^d$, Christian Schwinn${}^e$}
\vskip0.3cm
\begin{center}
{\em\footnotesize ${}^a$ Physik Department T31, Technische Universit\"at
      M\"unchen, James-Franck-Stra{\ss}e 1,\\[-0.05cm]
 D-85748 Garching, Germany}\\[-0.05cm]
{\em\footnotesize ${}^b$ Institut f\"ur Theoretische Physik E,
      RWTH Aachen University, D-52056 Aachen, Germany}\\[-0.05cm]
{\em\footnotesize ${}^c$ Institute for Theoretical Physics and Spinoza Institute, Utrecht
      University, 3508 TD Utrecht, The Netherlands}\\[-0.05cm]
{\em\footnotesize ${}^d$ Cavendish Laboratory, JJ Thomson Avenue,
      Cambridge CB3 0HE, United Kingdom }\\[-0.05cm]
{\em\footnotesize ${}^e$ Albert-Ludwigs Universit\"at Freiburg,
      Physikalisches Institut, D-79104 Freiburg, Germany}\\
\end{center}
\vskip0.9cm

\noindent
Ref.~\cite{Baernreuther:2013caa2} noted that there is an additional 
contribution to the singular terms in the threshold expansion of 
the heavy-quark pair production cross section at 
${\cal O}(\alpha_s^4)$ from the 
one-particle reducible vacuum polarization contribution 
\begin{displaymath}
\includegraphics[height=2.3cm]{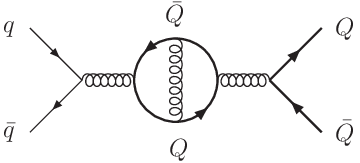}
\end{displaymath}
to the 2-loop virtual correction, 
which was omitted in our calculation. This affects only the 
quark-antiquark production channel $q\bar q \to Q\bar Q$. In 
consequence the expression
\[
\sigma^{(2,0)}_{q\bar{q},\bf{8}}\Big|_{\rm VP}
= \left(\cf - \frac{\ca}{2} \right) 8\pi^2 \ln \beta 
= -13.160 \ln\beta 
\]
must be added to Eq.~(5), which changes the coefficient of the 
$\ln\beta$ term in Eq.~(8) from 528.557 to 515.397. Eq.~(8) 
should read correctly
\begin{eqnarray*}
\label{eq:sigma-2-0-qq-numcorr}
\sigma^{(2)}_{q\bar{q}} &=&
\frac{3.60774}{\beta^2}
+\frac{1}{\beta}\Big(-140.368\lb2+32.106\lb1+3.95105\Big) \nonumber \\ &&
+910.222\lb4-1315.53\lb3+592.292\lb2+515.397\lb1+C^{(2)}_{q\bar q} \; .
\end{eqnarray*}
The new term amounts to a reduction of the coefficient of $\ln \beta$ 
by 2.5\%. The numerical effect on the total
cross section is negligible.

As discussed in Ref.~\cite{Beneke:2016kvz} in the more general context 
of pair production of squarks and gluinos, the above additional term 
should be interpreted as a correction to the ``Coulomb function'' 
$J_{R_\alpha}(E)$ in Eq.~(1) from an annihilation 
contribution, here from $Q\bar Q\to Q\bar Q$, to the NRQCD Lagrangian. 
Accordingly, the term 
\[
\nu^{R_\alpha,S}_{\rm ann} (-D_{R_\alpha})\,8\pi^2\ln\beta
\]
should be added to the general result for $\sigma_X^{(2)}$ in 
Eq.~(A.1). For a given heavy-particle state, $\nu^{R_\alpha,S}_{\rm ann}$ 
depends on the colour representation $R_\alpha$ 
and total spin $S$ of the pair. 
For $Q\bar Q$, only $\nu^{{\bf 8},1}_{\rm ann} = 2T_F=1$ is different 
from zero. Results for $\nu^{R_\alpha,S}_{\rm ann}$ as well as 
the quantity $v_{\rm spin}$, which appears in Eq.~(A.1), 
can be found in Ref.~\cite{Beneke:2016kvz} for the case of pair production 
of squarks and gluinos.

\end{document}